\documentclass{sigchi-ext}
\usepackage[T1]{fontenc}
\usepackage{textcomp}
\usepackage[scaled=.92]{helvet} 
\usepackage{graphicx} 
\usepackage{balance}  
\usepackage{booktabs} 
\usepackage{ccicons}  
\usepackage{ragged2e} 
\usepackage{xcolor}

\def\plaintitle{Generating Process-Centric Explanations to Enable Contestability in Algorithmic Decision Making: Challenges and Opportunities} 
\def\emptyauthor{}
\def\plainkeywords{process-centric explanations, contestability by design, reflexivity, discretionary choices, fairness perceptions}

\title{Generating \textit{Process-Centric Explanations} to Enable Contestability in Algorithmic Decision Making: Challenges and Opportunities}

\numberofauthors{3}

\author{%
  \alignauthor{%
    \textbf{Mireia Yurrita}\\
    \affaddr{Delft University of Technology} \\
    \affaddr{Delft, The Netherlands} \\
    \email{m.yurritasemperena@tudelft.nl} }\vfil \alignauthor{%
    \textbf{Agathe Balayn}\\
    \affaddr{Delft University of Technology}\\
    \affaddr{Delft, The Netherlands}\\
    \email{a.m.a.balayn@tudelft.nl} } \vfil \alignauthor{%
    \textbf{Ujwal Gadiraju}\\
    \affaddr{Delft University of Technology}\\
    \affaddr{Delft, The Netherlands}\\
    \email{u.k.gadiraju@tudelft.nl} } }

\definecolor{linkColor}{RGB}{6,125,233}
\hypersetup{%
  pdftitle={\plaintitle},
  pdfauthor={\emptyauthor},
  pdfkeywords={\plainkeywords},
  bookmarksnumbered,
  pdfstartview={FitH},
  colorlinks,
  citecolor=black,
  filecolor=black,
  linkcolor=black,
  urlcolor=linkColor,
  breaklinks=true,
}

\begin{document}

\CopyrightYear{2023}
\setcopyright{rightsretained}
\conferenceinfo{CHI'23,}{April  23--28, 2023, Hamburg, Germany}
\isbn{978-1-4503-6819-3/20/04}
\doi{https://doi.org/10.1145/3334480.XXXXXXX}

\copyrightinfo{\acmcopyright}

\maketitle

\RaggedRight{}

\begin{abstract}
Human-AI decision making is becoming increasingly ubiquitous, and explanations have been proposed to facilitate better Human-AI interactions. Recent research has investigated the positive impact of  explanations on decision subjects' fairness perceptions in algorithmic decision making. Despite these advances, most studies have captured the effect of explanations in isolation, considering explanations as ends in themselves, and reducing them to technical solutions provided through XAI methodologies. In this vision paper, we argue that the effect of explanations on fairness perceptions should rather be captured in relation to decision subjects' right to \textit{contest} such decisions. Since contestable AI systems are open to human intervention throughout their lifecycle, \textit{contestability} requires explanations that go beyond outcomes and also capture the rationales that led to the development and deployment of the algorithmic system in the first place. We refer to such explanations as \textit{process-centric explanations}. In this work, we introduce the notion of process-centric explanations and describe some of the main challenges and research opportunities for generating and evaluating such explanations. 
\end{abstract}


\printccsdesc

\section{Introduction}\label{sec:introduction}

Algorithmic systems are increasingly being used in \textit{high-risk} decision-making processes (e.g., in processes that grant access to education, employment and public services benefits~\cite{euaiact_2021}). These systems are claimed to have the potential to 
bring a wide-range of socio-economic benefits~\cite{floridi2019}. Yet, multiple concerns have been raised around the way in which algorithmic systems could 
lead to discrimination~\cite{oneil2016, buolamwini2018}. As a consequence, recent work has developed fairness-aware algorithmic systems to ensure the equity in the distribution of outcomes (i.e., \textit{distributive justice})~\cite{dwork2012, hardt2016}. However, even if algorithmic systems are fair according to some normative standard, they might not be \textit{perceived} as fair. 
Literature in legal and organizational psychology in human decision-making has long shown that decision subjects do not only care about the distribution of outcomes, but also care about the processes that led to those outcomes (i.e., \textit{procedural justice})~\cite{thibaut1975, leventhal1980}. Given the role that explanations play in unveiling \textit{how} an algorithmic decision was made, several studies 
\cite{dodge2019, binns2018} 
in HCI 
have looked into the effect of explanations on decision subjects' fairness perceptions.

\marginpar{%
  \vspace{-155pt} \fbox{%
    \begin{minipage}{0.925\marginparwidth}
     \vspace{1pc} \textbf{Justice } \\
       Multi-dimensional construct that studies \textit{fairness perceptions} across each of its dimensions~\cite{cropanzano2012}. \\
      \vspace{1pc} \textbf{Procedural justice } \\
       Justice dimension that aims to capture fairness perceptions regarding the process of a decision (i.e, \textit{procedural fairness perceptions)}~\cite{cropanzano2012}. \\
      \vspace{1pc} \textbf{Post-hoc contestability } \\
      Recourse and appeal processes~\cite{lyons2021}. \\
      \vspace{1pc} \textbf{Contestability by design}  \\
       Debates around the choices made throughout the development and deployment processes of algorithmic systems~\cite{alfrink2022_2}
    \end{minipage}}\label{sec:sidebar} }

Despite making important contributions, exisiting research has failed to overcome two main limitations. First, 
if explanations are not accompanied by an option to contest the decision, they may not contribute to effectively lending a voice to decision subjects --- a key aspect of procedural justice as defined by Thibaut and Walker~\cite{thibaut1975}. As it has been previously pointed out by legal scholars in discussions around the existence of \textit{the right for explanations} in the \textit{European Union's GDPR} 
(e.g., \cite{sarra2020, selbst2017}), explanations should be studied in relation to decision subjects' ability to contest automated decisions. \textit{Contestability} has been recognized as a form of procedural justice that increases perceptions of fairness~\cite{alfrink2022_2}. In turn, contestability requires explanations so that decision subjects can put forward solid arguments~\cite{sarra2020}. 

Second, explanations have  been restricted to outcome explanations enabled by technical eXplainable AI (XAI) methods. However, subjective discretionary choices made throughout the development and deployment of algorithmic systems affect the resulting algorithmic behaviour~\cite{cambo2022}, which decision subjects might want to contest~\cite{alfrink2022_2}.
It is, therefore, necessary to extend explanations to the system's entire lifecycle and go from \textit{post-hoc contestability} 
to \textit{contestability by design}~\cite{alfrink2022_2} 
.

In this vision paper, we argue that the effect of explanations on decision subjects' fairness perceptions should be studied in relation to their ability to contest automated decisions. We introduce and conceptualize \textit{process-centric explanations} as necessary elements to enable contestability by design~\cite{alfrink2022_2}. Process-centric explanations capture the decisions made throughout the development and deployment of algorithmic systems (e.g., information about training data). However, as opposed to concepts such as \textit{data-centric explanations}~\cite{anik2021}, process-centric explanations also distil the rationales behind those choices to enable open scrutiny. In the following sections, we outline the need for and nature of \textit{process-centric explanations} and outline the challenges and research opportunities for generating these explanations. We extend the conceptual framework presented by Schoeffer et al. \cite{schoeffer2022_2} at the ACM CHI 2022 Workshop on Human-Centered Explainable AI (HCXAI) by incorporating process-centric explanations as key elements in the interplay between procedural fairness, reliance, and distributive fairness (Figure \ref{fig:framework} in the appendix). This framework is complementary to the framework on contestability by design suggested by Alfrink et al.~\cite{alfrink2022_2}. 

\section{Process-centric Explanations} \label{sec:conceptualframework}

\subsection{Need for process-centric explanations.}
In practice and use, algorithmic systems are often regarded as ``objective'' means for decision making. However, there are several ``undisclosed yet impactful''~\cite{cambo2022} subjective discretionary choices that different stakeholders make throughout the development and deployment of machine learning (ML) systems~\cite{passi2018, passi2020, sambasivan2021}. For example, to translate highly complex strategic business goals into design requirements that represent tractable engineering problems, appropriate target variables or proxies to those variables need to be identified~\cite{passi2019}. In relation to data, various values and decisions shape the way data is collected (e.g., decisions around which data is included, which data is excluded and what the measurement of good data is)~\cite{pine2015}, transformed (e.g., implications of removing data from minority classes in the form of outliers)~\cite{balayn2021_2}, analyzed~\cite{muller2019} and visualized~\cite{kandel2012}. 
The subjectivity ingrained in these activities~\cite{tubella2022, thakkar2022} has often been identified as the root cause behind  harmful downstream impacts~\cite{sambasivan2021}. Process-centric explanations rely on \textit{reflexivity} 
to turn 
normative assumptions into \textit{meaningful information}~\cite{sarra2020} that facilitate \textit{contestability by design}~\cite{alfrink2022_2}.

\marginpar{%
  \vspace{-105pt} \fbox{%
    \begin{minipage}{0.925\marginparwidth}
     \vspace{1pc} \textbf{Reflexivity} \\
        Practice to inspect the presumptions and every-day processes taken from granted among Machine Learning practitioners (e.g., data engineers and data scientists)~\cite{miceli2021_2, hirsbrunner2022, cambo2022, elish2018}. \\
      \vspace{1pc} \textbf{Collaborative reflexive practices } \\
      Processes where struggles between stakeholders and disciplines are made productive for reflection~\cite{hirsbrunner2022}.\\
    \end{minipage}}\label{sec:sidebar} }

\subsection{Characteristics of process-centric explanations.}
Process-centric explanations take inspiration from and expand constructs such as \textit{data-centric explanations}~\cite{anik2021}, \textit{Datasheets}~\cite{gebru2021}, \textit{Dataset Nutrition Labels}~\cite{holland2018} or \textit{Model Cards}~\cite{mitchell2019}. Data-centric explanations include information about training data: mechanisms for data collection (i.e., amount of data collected, source of the data, collector of the data and labeling process), demographics (i.e., information on distribution of gender, race, age, and country of data instances), recommended usage, and potential issues. 
Similar to data-centric explanations, process-centric explanations go beyond outcome transparency and unpack details about the process that led to the development of the algorithmic system in question. However, we view process-centric explanations as a means towards empowering decision subjects to exercise their right to contest an automated decision. We echo Selbst et al. \cite{selbst2017} and advocate for functional explanations that enable a decision subject to take action. In practical terms, as opposed to data-centric explanations, this means that process-centric explanations focus on unveiling the rationales behind the discretionary choices made throughout the ML system lifecycle. For instance, instead of simply describing the way in which data was collected (data-centric explanation), a process-centric explanation would outline the reasons why a certain source of data was selected and a reflection around the potential harms that this specific choice could lead to (e.g., reflections around the rationales for not including some sub-population in the data~\cite{muller2022}). Making these rationales explicit complement the technical description (``how'') with a justification (``why'')~\cite{sarra2020, henin2021}  and set the grounds for a debate around data and modeling choices~\cite{alfrink2022_2}. 
As opposed to Datasheets~\cite{gebru2021}, Dataset Nutrition Labels~\cite{holland2018} and Model Cards~\cite{mitchell2019}, the target users of process-centric explanations are decision subjects (or their legal representatives~\cite{lyons2021}) and not dataset or model creators and consumers. That is to say, process-centric explanations should equip decision subjects with a ``higher or more specific level information''~\cite{sarra2020} that will enable them to have a rational and fruitful dialectical exchange as part of the contestation process. For instance, these include decisions that reflect the values of an organization (e.g., basic choices that made an organization prioritize an opaque system in favor of accuracy).

\subsection{Generating and evaluating process-centric explanations: \\Challenges and opportunities.}
There are several challenges and concomitant research opportunities related to effectively fostering reflexive practices throughout the ML pipeline and generating process-centric explanations. We identify and synthesize these challenges and opportunities across three main levels of abstraction
. 

Challenges on a \textit{micro level} (\textbf{individuals}). One of the main challenges in fostering reflexive practices among ML practitioners is raising awareness around the subjective nature of discretionary choices and the downstream impacts these might have. For instance, during data wrangling, curation, and feature engineering data scientists might be making decisions on how to treat the data which can, in turn, introduce biases leading to discrimination~\cite{muller2022}. Challenges at this level of abstraction include (1) technical development of adequate infrastructure as well as (2) further training and guidance for ML practitioners to be able to adopt reflexive practices. As far as technical infrastructure is concerned, previous work has pinpointed the need to develop technical interfaces for versioning data~\cite{schelter2019} or for keeping track of how data transformations might affect fairness metrics~\cite{balayn2021_2}. When it comes to training practitioners in reflexive practices, studies exploring reflexive tools like fairness toolkits, have raised concerns about the lack of actionable guidance~\cite{deng2022} and the need to further support practitioners on reflecting around the downstream harms that their discretionary choices might lead to~\cite{madaio2020, madaio2022}. This includes providing resources for practicing reflection through real-world examples~\cite{deng2022, holstein2019, lee2021_2} and onboarding material~\cite{lee2021_2}, or integrating more targeted guidance in the form of wizards within the reflexive tools~\cite{balayn2021_2}. It should be noted that, even when guidance in the form of Datasheets~\cite{gebru2021} is given, the connection between these reflexive documents and responsible AI practices is not straightforward for many~\cite{heger2022}. It is, therefore, necessary to make these relations explicit.

Challenges on a \textit{mezzo level} (\textbf{negotiations}). How algorithmic systems work is not a given but is rather negotiated in the problem formulation stage~\cite{passi2020}. Similarly, how  a problem is specified and operationalized is always negotiated~\cite{muller2022}. Given that negotiation processes take place throughout the ML pipeline, previous work has identified the need to promote \textit{collaborative reflexive practices}
. Collaborative reflexive practices require, in turn, tools to facilitate communication~\cite{deng2022, holstein2019, rakova2021} and digital objects (e.g., data visualizations and reports tailored for non-engineers~\cite{deng2022, lee2021_2}) to enable discursive negotiation~\cite{hirsbrunner2022}.

Challenges on a \textit{macro level} (\textbf{governance}). Initiatives and policies at an organization level are key to effectively fostering reflexive practices among practitioners and enabling the generation of process-centric explanations. Previous work has identified the need to account for business imperatives~\cite{madaio2022} and organizational constraints~\cite{green2021_2} when asking individuals and teams to engage with reflexive practices. The importance of implementing organization-wide strategies (e.g., resources~\cite{madaio2022} and support~\cite{madaio2020}) can, therefore, not be ignored. Such strategies include, but are not limited to, generating shared understandings of quality objectives among stakeholders, improving cross-organizational documentation and creating incentives for stakeholders to engage in reflexive practices~\cite{thakkar2022}.

Process-centric explanations can be generated by documenting the rationales distilled from these reflexive practices. Going from documentation to process-centric explanations results in some additional challenges. First, there is a need to define what can be contested~\cite{lyons2021} and therefore tailor process-centric explanations to information relevant for contestability. Second, this information should be presented to cater to the needs of decision subjects (or their legal representatives~\cite{lyons2021}). 
Third, evaluation protocols should also be created to ensure the effectiveness of process-centric explanations in enacting contestability. We need to explore how to effectively and holistically evaluate process-centric explanations across different levels. 
These evaluation protocols might require, for instance, to iteratively update process-centric explanations. The way in which such protocols can be orchestrated and operationalized represents a promising future research direction.

\section{Acknowledgments}\label{sec:acks}

This work was partially supported by the European Union’s Horizon 2020 research and innovation programme under the Marie Skłodowska‐Curie grant agreement No 955990 and the TU Delft AI Initiative.




\bibliographystyle{SIGCHI-Reference-Format}
\bibliography{references}

\newpage
\appendix
\onecolumn

\begin{figure*}
\centering
  \includegraphics[scale = 0.115]{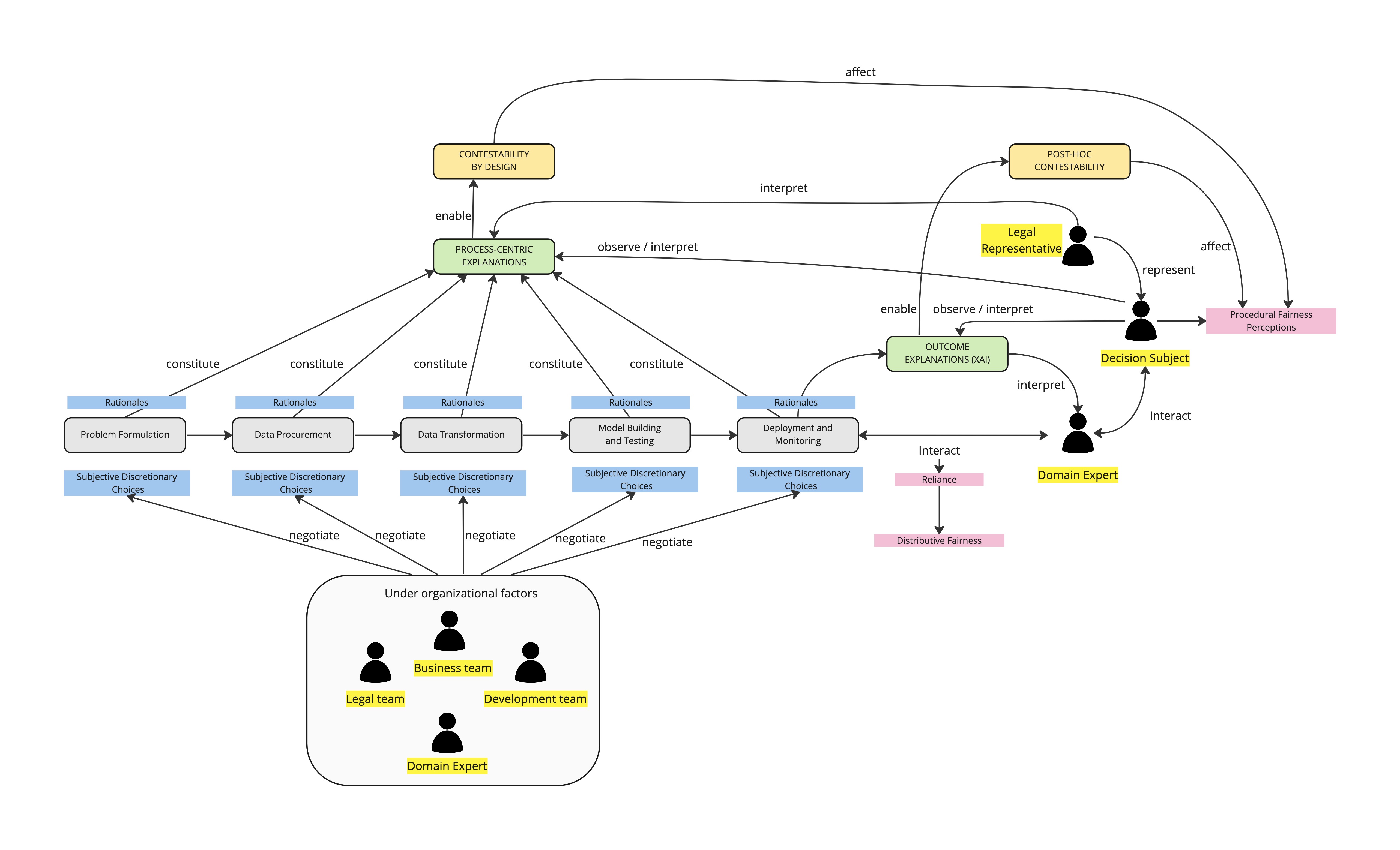}
  \caption{Overview of the conceptual framework that locates process-centric explanations as necessary elements for contestability by design, which in turn affects decision subjects fairness perceptions. In pink, we highlight the concepts defined by Schoeffer et al. \protect \cite{schoeffer2022_2}. Grey refers to different stages of the ML pipeline, green refers to explanatory elements, orange refers to contestability elements, yellow refers to different stakeholders, and blue refers to discretionary choices and their rationales.}
  \label{fig:framework}
\end{figure*}

\end{document}